
\frenchspacing

\parindent15pt
\overfullrule=0pt
\abovedisplayskip4pt plus2pt
\belowdisplayskip4pt plus2pt 
\abovedisplayshortskip2pt plus2pt 
\belowdisplayshortskip2pt plus2pt  

\font\twbf=cmbx10 at12pt
 at12pt
 at12pt

\font\ninerm=cmr9 
\font\nineit=cmti9 
\font\ninesy=cmsy9 
\font\ninei=cmmi9 
\font\ninebf=cmbx9 

\font\sevenrm=cmr7  
 
\font\seveni=cmmi7  
\font\sevensy=cmsy7 

\font\fivenrm=cmr5  
\font\fiveni=cmmi5  
\font\fivensy=cmsy5 

\def\nine{%
\textfont0=\ninerm \scriptfont0=\sevenrm \scriptscriptfont0=\fivenrm
\textfont1=\ninei \scriptfont1=\seveni \scriptscriptfont1=\fiveni
\textfont2=\ninesy \scriptfont2=\sevensy \scriptscriptfont2=\fivensy
\textfont3=\tenex \scriptfont3=\tenex \scriptscriptfont3=\tenex
\def\rm{\fam0\ninerm}%
\textfont\itfam=\nineit    
\def\it{\fam\itfam\nineit}%
\textfont\bffam=\ninebf 
\def\bf{\fam\bffam\ninebf}%
\normalbaselineskip=11pt
\setbox\strutbox=\hbox{\vrule height8pt depth3pt width0pt}%
\normalbaselines\rm}

\hsize30cc
\vsize44cc
\nopagenumbers

\def\luz#1{\luzno#1?}
\def\luzno#1{\ifx#1?\let\next=\relax\yyy
\else \let\next=\luzno#1\xxx\fi\next}
\def\sp#1{\def\xxx{\kern1.7pt}\def\yyy{\kern-1.7pt}\luz{#1}}
\def\spa#1{\def\xxx{\kern1pt}\def\yyy{\kern-1pt}\luz{#1}}

\newcount\beg
\newbox\aabox
\newbox\atbox
\newbox\fpbox
\def\abbrevauthors#1{\setbox\aabox=\hbox{\sevenrm\uppercase{#1}}}
\def\abbrevtitle#1{\setbox\atbox=\hbox{\sevenrm\uppercase{#1}}}
\long\def\pag{\beg=\pageno
\def\leftheadline{\noindent\rlap{\nine\folio}\hfil\copy\aabox\hfil}
\def\rightheadline{\noindent\hfill\copy\atbox\hfill\llap{\nine\folio}}
\def\phead{
\noindent\vbox{
\centerline{\hfill LMU-TPW 96-5\ }

\centerline{\hfill February 2, 1996\ }}}
\footline{\ifnum\beg=\pageno \hfill\nine[\folio]\hfill\fi}
\headline{\ifnum\beg=\pageno\phead
\else
\ifodd\pageno\rightheadline \else \leftheadline \fi 
\fi}}

\newbox\tbox
\newbox\aubox
\newbox\adbox
\newbox\mathbox

\def\title#1{\setbox\tbox=\hbox{\let\\=\cr 
\baselineskip14pt\vbox{\twbf\tabskip 0pt plus15cc
\halign to\hsize{\hfil\ignorespaces \uppercase{##}\hfil\cr#1\cr}}}}

\newbox\abbox
\setbox\abbox=\vbox{\vglue18pt}

\def\author#1{\setbox\aubox=\hbox{\let\\=\cr 
\nine\baselineskip12pt\vbox{\tabskip 0pt plus15cc
\halign to\hsize{\hfil\ignorespaces \uppercase{\spa{##}}\hfil\cr#1\cr}}}%
\global\setbox\abbox=\vbox{\unvbox\abbox\box\aubox\vskip8pt}}

\def\address#1{\setbox\adbox=\hbox{\let\\=\cr 
\nine\baselineskip12pt\vbox{\it\tabskip 0pt plus15cc
\halign to\hsize{\hfil\ignorespaces {##}\hfil\cr#1\cr}}}%
\global\setbox\abbox=\vbox{\unvbox\abbox\box\adbox\vskip16pt}}

\def\mathclass#1{\setbox\mathbox=\hbox{\footnote{}{1991 {\it Mathematics 
Subject Classification}\/: #1}}}

\long\def\maketitlebcp{\pag\unhbox\mathbox
\vglue7cc
\box\tbox
\box\abbox
\vskip8pt}

\long\def\abstract#1{{\nine{\bf Abstract.} 
#1

}}

\def\section#1{\vskip-\lastskip\vskip12pt plus2pt minus2pt
{\bf #1}}

\def\endproof{\nobreak\kern5pt\nobreak\vrule height4pt width4pt depth0pt
\vskip4pt plus2pt}

\newbox\refbox
\newdimen\refwidth
\long\def\references#1#2{{\nine
\setbox\refbox=\hbox{\nine[#1]}\refwidth\wd\refbox\advance\refwidth by 12pt%
\def\textindent##1{\indent\llap{##1\hskip12pt}\ignorespaces}
\vskip24pt plus4pt minus4pt
\centerline{\bf References}
\vskip12pt plus2pt minus2pt
\parindent=\refwidth
#2

}}

\def\footnoterule{\kern -3pt \hrule width 4cc \kern 2.6pt}

\catcode`@=11
\def\vfootnote#1%
{\insert\footins\bgroup\nine\interlinepenalty\interfootnotelinepenalty%
\splittopskip\ht\strutbox\splitmaxdepth\dp\strutbox\floatingpenalty\@MM%
\leftskip\z@skip\rightskip\z@skip\spaceskip\z@skip\xspaceskip\z@skip%
\textindent{#1}\footstrut\futurelet\next\fo@t}
\catcode`@=12

\mathclass{Primary 81R50; Secondary 17B37.}

\abbrevauthors{G. Fiore and P. Schupp}
\abbrevtitle{Statistics and Quantum Groups...}

\title{Statistics and Quantum Group Symmetries}
\author{Gaetano\ Fiore}
\address{Sektion Physik Universit\"at M\"unchen, LS Prof. Wess, \\
Theresienstr. 37, D-80333 M\"unchen, Germany\\
E-mail: Gaetano.Fiore@physik.uni-muenchen.de, gfiore@na.infn.it}

\author{Peter\ Schupp}
\address{Sektion Physik Universit\"at M\"unchen, LS Prof. Wess, \\
Theresienstr. 37, D-80333 M\"unchen, Germany\\
E-mail: Peter.Schupp@physik.uni-muenchen.de}

\maketitlebcp

\footnote{}{First author supported by a postdoc fellowship
of the A. v. Humboldt Foundation.}

\abstract{Using ``twisted'' realizations of the symmetric groups,
we show that Bose and Fermi statistics are compatible with transformations
generated by compact quantum groups of Drinfel'd type.}

\section{1. Introduction.} 

In recent times quantum groups have been
often
suggested as candidates for generalized
symmetry transformations in Quantum Field 
Theory. One way to reach QFT is to  
deform canonical commutation relations
of fields in such a way that they become quantum
group covariant [1]. Another way is 
first to find  a consistent procedure
to implement quantum group transformations in Quantum
Mechanics with a {\it finite} number of identical 
particles and then to pass to QFT through second quantization;
in this approach the ``particle interpretation'' of QFT is
the starting point. In the present work we adopt this
second approach and give a consistent first quantized
construction [2].

Are the notions of {\bf identical particles} and
{\bf quantum (group) symmetry} compatible?
In other words, can the Hilbert space
of states of $n$ bosons/fermions carry both 
a representation of the symmetric group $S_n$
and of $H$?---One might think that this is impossible.

Let $H$ be some $*$ Hopf algebra and
$\rho: H \rightarrow {\cal A}$ 
a realization of $H$ in a $*$-algebra $\cal A$
of operators that act on a one-particle Hilbert space ${\cal H}$. We say:
``the system transforms under the action of $H$''.
If $H$  is not co-commutative, then the
action of $H$ on ${\cal H} \otimes {\cal H}$, defined by 
$\rho^{(2)} = (\rho \otimes
\rho)\circ\Delta$, where $\Delta$ is the coproduct,
does not preserve but rather {\it mixes} the symmetric  and
antisymmetric subspaces
$({\cal H} \otimes {\cal H})_\pm$ defined by 
$P_{12} ({\cal H} \otimes {\cal H})_\pm =
\pm ({\cal H} \otimes {\cal H})_\pm$ respectively ($P_{12}$ 
denotes the permutation
operator). Fermions and bosons in the
ordinary sense are impossible. 

This is true even if $H$ is just a {\it slight} deformation of
a co-commutative Hopf algebra, because
a drastic and unacceptable discontinuity of
the number of allowed states of the multi-particle system 
would appear in the limit of vanishing deformation parameter.

We want to show that there exists a way out,
at least if $H$ is a compact section of a quantum group 
of Drinfel'd type, {\it i.e.} a quantization of a Poisson-Lie group.

The apparent incompatibility arises from two facts:
1)  in the (ordinary) formulation of Quantum Mechanics
one associates to each {\it separate} tensor factor one
of the particles; 2) the coproduct does not
``treat'' different tensor factors symmetrically.
In the following we want to show that point 1)  is only a possibility, 
not a necessity.

\section{2. Twisted symmetrization postulates}

The ``permutation group" $S_n$ enters the definitions of the
Hilbert space and the algebra of operators of a system of 
bosons/fermions. The first step is to show that the elements of 
$S_n$ can be realized in a way other than by ordinary 
permutators.

The standard symmetry postulates characterizing a system of {\bf 2}
bosons or fermions are:
$$\eqalignno{
& P_{12}|u\rangle_\pm = \pm|u\rangle_\pm \hskip.9cm \hbox{for} 
\hskip.3cm  |u\rangle_{\pm} \in
({\cal H} \otimes {\cal H})_\pm & (1)  \cr
& a:({\cal H} \otimes {\cal H})_\pm \rightarrow ({\cal H} \otimes 
{\cal H})_\pm \hskip.9cm \hbox{for}
\hskip.3cm a \in (\hbox{$\cal A$} \otimes \hbox{$\cal A$})_+ & (2) \cr
& *_2: (\hbox{$\cal A$} \otimes \hbox{$\cal A$})_+\rightarrow 
(\hbox{$\cal A$} \otimes \hbox{$\cal A$})_+, & (3) \cr}
$$%
where $*_2 \equiv * \otimes *$ and 
$ (\hbox{$\cal A$} \otimes \hbox{$\cal A$})_+ := \{ a \in \hbox{$\cal A$} 
\otimes \hbox{$\cal A$} : [P_{12}, a ] = 0  \}.$
Equation (1) defines bosonic $(+)$ and fermionic $(-)$
states.
Equation (2) follows from
$[P_{12}, (\hbox{$\cal A$}\otimes\hbox{$\cal A$})_+] = 0$
and shows that symmetrized operators map  boson (fermion) 
states into boson (fermion) states.

For a unitary (and in general not symmetric)
operator $F_{12} \in \hbox{$\cal A$} \otimes \hbox{$\cal A$}$, 
$F_{12}^{*_2} = F_{12}^{-1}$ we define
$$\eqalignno{
({\cal H} \otimes {\cal H})_{\pm}^{{F_{12}}} & := F_{12} 
({\cal H} \otimes {\cal H})_\pm & (4) \cr
P_{12}^{F_{12}}& :=  F_{12} P_{12} F_{12}^{-1} & (5) \cr
(\hbox{$\cal A$} \otimes \hbox{$\cal A$})_+^{F_{12}}& :=  
F_{12}(\hbox{$\cal A$} \otimes \hbox{$\cal A$})_+ F_{12}^{-1} & ~ \cr
\qquad & =  \{ a \in \hbox{$\cal A$} \otimes \hbox{$\cal A$} : 
[P_{12}^{F_{12}}, a ] = 0  \} & (6) \cr
}$$
with $(\hbox{$\cal A$} \otimes \hbox{$\cal A$})_+$ as defined above. Note that 
$\big(P_{12}^{F_{12}}\big)^2=\hbox{id}=\big(P_{12}\big)^2$.
We find in complete analogy to equations (1) -- (3)
$$\eqalignno{
& P_{12}^{F_{12}}|u\rangle_\pm = \pm|u\rangle_\pm \hskip.9cm 
\hbox{for} \hskip.3cm
|u\rangle_\pm \in
({\cal H} \otimes{\cal H})_\pm^{F_{12}} & (7) \cr
& a:({\cal H} \otimes {\cal H})^{F_{12}}_\pm \rightarrow ({\cal H} 
\otimes {\cal H})^
{F_{12}}_\pm \hskip.9cm \hbox{for}\hskip.3cm a \in (\hbox{$\cal A$} \otimes
\hbox{$\cal A$})_+^{F_{12}}  &  (8) \cr
& *_2: (\hbox{$\cal A$} \otimes \hbox{$\cal A$})_+^{F_{12}}\rightarrow 
(\hbox{$\cal A$} \otimes \hbox{$\cal A$})_+^{F_{12}}  
& (9) \cr 
}$$
and $a^{F_{12}}:= F_{12} a F_{12}^{-1}$ is hermitean iff $a$ is. Eq.'s
(7) -- (9) are formally identical to eq.'s
(1) -- (3) and provide an alternative quantum mechanical description 
of a system of two bosons/fermions. The unitarity of $F_{12}$ 
guarantees that from the kinematical viewpoint the new ``twisted'' description
is physically equivalent to the standard one.

The generalization to {\bf n} bosons or fermions is straightforward:
Let $F_{12\ldots n} \in \hbox{$\cal A$}^{\otimes n}$
be unitary,
{\it i.e.} $(F_{12\ldots n})^{*_n} = (F_{12\ldots n})^{-1}$, where $*_n :=
*^{\otimes n}$, and define
$$\eqalignno{
({\cal H} \otimes\ldots\otimes {\cal H})_\pm^{F_{12\ldots n}}& :=  
F_{12\ldots n}
({\cal H} \otimes\ldots\otimes {\cal H})_\pm & (10) \cr
P_{12}^{F_{12\ldots n}}& :=  F_{12\ldots n} P_{12}
(F_{12\ldots n})^{-1} \cr
& \vdots & (11)  \cr
P_{n-1,n}^{F_{12\ldots n}}& :=  F_{12\ldots n} P_{n-1,n}
(F_{12\ldots n})^{-1}\cr
(\hbox{$\cal A$} \otimes\ldots\otimes\hbox{$\cal A$})_+^{F_{12\ldots n}}& 
:=  F_{12\ldots n}
(\hbox{$\cal A$}\otimes\ldots\otimes\hbox{$\cal A$})_+(F_{12\ldots n})^{-1} 
& (12) \cr
}$$
with $(\hbox{$\cal A$}\otimes\ldots\otimes\hbox{$\cal A$})_+ 
:= \{ a \in \hbox{$\cal A$}\otimes\ldots\otimes\hbox{$\cal A$} : 
[P_{i,i+1} , a] =0 ,i=1,\ldots n-1\},$
and $P_{i,i+1}$ the permutator of the $i^{th},(i\!+\!1)^{th}$
tensor factors.
Then
$$\eqalignno{
&P_{i,i+1}^{F_{12\ldots n}}|u\rangle_\pm =
\pm |u\rangle_\pm \hskip.9cm\hbox{for}\hskip.3cm |u\rangle_\pm
\in ({\cal H} \otimes\ldots\otimes
{\cal H})_\pm^{F_{12\ldots n}} & (13)  \cr
&a: ({\cal H} \otimes\ldots\otimes {\cal H})_\pm^{F_{12\ldots n}}\rightarrow
({\cal H} \otimes\ldots\otimes {\cal H})_\pm^{F_{12\ldots n}} & (14) \cr
& \hskip.9cm\hskip.9cm\hskip.9cm\hskip.9cm\hskip.9cm\hbox{for}\hskip.3cm a \in
(\hbox{$\cal A$} \otimes\ldots\otimes\hbox{$\cal A$})_+^{F_{12\ldots n}}\cr
&*_n: (\hbox{$\cal A$} \otimes \ldots \otimes
\hbox{$\cal A$})_+^{F_{12\ldots n}}\rightarrow
(\hbox{$\cal A$} \otimes \ldots \otimes\hbox{$\cal A$})_
+^{F_{12\ldots n}}. & (15) \cr
}$$
Equation (14) follows from $[P_{i,i+1}^{F_{12\ldots n}}, 
(\hbox{$\cal A$}\otimes\ldots\otimes\hbox{$\cal A$})_+^ {F_{12\ldots n}}] = 0.$

The operators $P_{i,j}^{F_{12\ldots n}}$ satisfy the same 
algebraic relations as the  $P_{i,j}$. They therefore provide a 
different realization of the group $S_n$. Eq.'s  (13) -- (15) 
give an alternative quantum mechanical description 
of a system of $n$ bosons/fermions;
note that in the latter equations 
the twist $F_{12\ldots n}$ does not  
appear explicitly any more.

\section{3. Identical versus distinct particles }

It is every day's experience that
in many situations {\it identical} particles can be equivalently
treated as {\it distinct}. If this were not the case,
one could not describe some particles (in our ``laboratory'', say) 
without describing
(with a unique huge wave-function)
all the other particles of the same kind in the universe.

There exists a precise correspondence 
between these two descriptions in the standard formalism;
in the twisted approach
the twist $F$ directly enters the rule governing this
correspondence.
      
Consider as gedanken experiment the scattering of
two identical particles. One can distinguish three stages:
In the initial stage 
the two particles are far apart, i.e. are prepared
in two separate orthonormal one-particle states $|\psi_1\rangle$,
$|\psi_2\rangle$. In the intermediate stage, the particles 
approach each other and scatter. In the
final stage, long after the collision, the particles are
again far apart and are detected by one-particle detectors.
Since the preparation and measurement
are essentially one-particle processes, we need to:
\item{(a)} Translate the initial data (two one-particle 
states $|\psi_1\rangle$, $|\psi_2\rangle$) into a properly
\hbox{(anti-)}\-sym\-metrized two-particle state $|\psi\rangle$; in fact, 
the collision is correctly described only if we apply a
symmetric evolution operator to such a state.
\item{(b)} Translate the final \hbox{(anti-)}symmetrized two-particle 
state $|\psi'\rangle$
into one-particle data, i.e. a superposition $|\psi_d'\rangle$ 
of (correlated) orthonormal one-particle states.

In the twisted case these translations are done according to
the following rule: $|\psi\rangle$ is obtained by
$$
|\psi\rangle:={F_{12}\over\sqrt{2}}\left(
|\psi_1\rangle\otimes|\psi_2\rangle\pm |\psi_2\rangle
\otimes|\psi_1\rangle\right)\in ({\cal H}\otimes{\cal H})^{F_{12}}_{\pm} 
\eqno{(16)}
$$
and, with the final state of the form
$|\psi'\rangle= \sum\limits_i,{F_{12}\over\sqrt{2}}\left(
|\psi^i_1\rangle\otimes|\psi^i_2\rangle\pm |\psi^i_2\rangle
\otimes|\psi^i_1\rangle\right)$,
where $\langle\psi^i_1|\psi^i_2\rangle=0$, 
$|\psi_d'\rangle$ is obtained as 
$$
|\psi_d'\rangle=
\sum\limits_i|\psi^i_1\rangle\otimes|\psi^i_2\rangle. \eqno{(17)}
$$

\section{4. Quantum Symmetries}

While their introduction was shown to be consistent,
there was so far no need for the $F_{12\ldots n}$. Now we take the
issue of
quantum group symmetries into consideration.
We require that multi-particle systems 
transform under Hopf algebra actions in a way
that is consistent with  
twisted (anti-)symmetric states and operators. 
The twist ${F_{12\ldots n}}$
is strictly related to the coproduct $\Delta$.
A trivial ${F_{12\ldots n}}$ is only possible for
a cocomutative $\Delta$.

\section{4.1 Transformation of states and operators}

The transformation $\triangleright$ of ${\cal H}$ under $H$ shall be
given by a $*$-algebra homomorphism
$\rho:H \rightarrow \hbox{$\cal A$}$
with $\rho(x)\rho(y) = \rho(x y)$, $\rho(x^*)
=\rho^\dagger(x)$, and $\rho(1) = {\bf 1}$:
$$
|\psi\rangle \rightarrow \rho(x) |\psi\rangle =: 
x \triangleright |\psi\rangle,\qquad |\psi\rangle \in
{\cal H},\quad x \in H. \eqno{(18)}
$$%
The corresponding transformation of operators under the action of $H$ is
$$
\hbox{$\cal O$} \rightarrow \rho(x_{(1)})  \hbox{$\cal O$}  
\rho(S x_{(2)}) =: x \triangleright \hbox{$\cal O$},\qquad
x_{(1)} \otimes x_{(2)} \equiv \Delta(x) \in H \otimes H, \quad 
\hbox{$\cal O$} \in \hbox{$\cal A$}. \eqno{(19)}
$$%

The transformation of multi particle systems, i.e. of $|\psi^{(n)}\rangle \in 
{\cal H}^{\otimes n}$ and $\hbox{$\cal O$}^{(n)} \in 
\hbox{$\cal A$}^{\otimes n}$ is obtained simply by
replacing $\rho$ in expressions (18) and (19) by 
$$
\rho^{(n)}= \rho^{\otimes n} \circ \Delta^{(n-1)}. \eqno{(20)}
$$%

The reader may wonder what happened to the ordinary commutator (i.e. the
adjoint action in the undeformed setting): The commutator is in general
distinct from the quantum adjoint action (19), but 
it still plays a role here: An operator $\hbox{$\cal O$}$ is
symmetric (invariant) under the transformations generated by $h \subset H$ if 
$x \triangleright \hbox{$\cal O$} = \epsilon(x) \hbox{$\cal O$}$ for all 
$x \in h$; \
it may be simultaneously diagonalizable with elements $x \in h$ if 
$[\rho(x) , \hbox{$\cal O$} ] = 0.$
Symmetry and vanishing commutator
coincide if $\Delta(h) \subset h \otimes H$.

The transformation of ${\cal H}$ and $\hbox{$\cal A$}$ under 
the Hopf algebra $H$ should be
consistent with the twisted symmetrization postulates of section~2: We 
hence require that
$$
H: \; ({\cal H}^{\otimes n})_\pm^{F_{12\ldots n}} \rightarrow 
({\cal H}^{\otimes n})_\pm^{F_{12\ldots n}},  \eqno{(21)}
$$
$$
H: \; (\hbox{$\cal A$}^{\otimes n})_+^{F_{12\ldots n}} \rightarrow 
(\hbox{$\cal A$}^{\otimes n})_+^{F_{12\ldots n}}     \eqno{(22)}
$$
for a suitably chosen $F_{12\ldots n}$.
The conditions to be satisfied are
$$
[\rho^{(n)}(x) , P^{F_{12\ldots n}}_{i,i+1} ] = 0, \qquad \forall x \in H,
\quad i = 1,\ldots,n-1, \eqno{(23)}
$$
because then the action of $H$ will commute with the twisted permutations
$P^{F_{12\ldots n}}_{i,i+1}$:
$$
P^F ( x \triangleright |\psi^{(n)}\rangle) = 
P^F \rho^{(n)}(x)|\psi^{(n)}\rangle
= \rho^{(n)}(x) ( P^F |\psi^{(n)}\rangle) = x \triangleright 
( P^F |\psi^{(n)}\rangle) \eqno{(24)}
$$
and 
$$\eqalignno{
[P^F , x \triangleright \hbox{$\cal O$}^{(n)} ]  &= [P^F,\rho^{(n)}(x_{(1)}) 
\hbox{$\cal O$}^{(n)}
\rho^{(n)}(S x_{(2)}) ] = \rho^{(n)}(x_{(1)})[P^F,\hbox{$\cal O$}^{(n)}] 
\rho^{(n)}(S x_{(2)}) & \cr
&= x \triangleright [P^F , \hbox{$\cal O$}^{(n)}] & (25)}
$$
(Here we have suppressed the indices on $P^F$.)

Conditions (23) are equivalent to requiring 
$$
\rho^{(n)}(H) \subset (\hbox{$\cal A$}^{\otimes n})_+^{F_{12\ldots n}}
\eqno{(26)}
$$%
for some $F_{12\ldots n}$. This is certainly satisfied if
$$
\rho^{(n)}(H) = F_{12..n} \rho^{(n)}_c(H) F_{12..n}^{-1},\eqno{(27)}
 $$
where $\rho^{(n)}_c:=\rho^{\otimes n} \circ \Delta_c^{(n-1)}$
and $\Delta_c$ is a cocommutative coproduct. The following theorem
will be our guidance in finding the right $F$'s when 
$H=U_q(g)$.

\section{4.2 Drinfel'd Proposition 3.16 in Ref. [3]}

{\it
\item{1.} There exists an algebra isomorphism
$\phi: U_q\hbox{\bf g} \tilde{\leftrightarrow}
(U \hbox{\bf g})([[h]])$, where $h = \ln q$ is the deformation parameter.
\item{2.} If we identify the isomorphic elements of $U_q\hbox{\bf g}$ and
$(U \hbox{\bf g})([[h]])$
then there exists an $\hbox{$\cal F$} \in U_q\hbox{\bf g} 
\otimes U_q\hbox{\bf g}$ such that:
$$
\Delta(a) = \hbox{$\cal F$} \Delta_c(a) \hbox{$\cal F$}^{-1},
\hskip.9cm \forall a \in U_q\hbox{\bf g} 
\tilde{=}
(U \hbox{\bf g})([[h]]) \eqno{(28)} 
$$%
where $\Delta$ is the coproduct of $U_q\hbox{\bf g}$ and $\Delta_c$ is the
(co-commutative) coproduct of $U(\hbox{\bf g})$.
\item{3.} $(U \hbox{\bf g})([[h]])$ is a quasi-triangular quasi-Hopf algebra
(QTQHA) with universal $\hbox{$\cal R$}_\Phi = q^{t/2}$ and a
quasi-coassociative structure given by an element $\Phi \in
\left((U \hbox{\bf g})^{\otimes 3}([[h]])\right)$ that is expressible in 
terms of $\hbox{$\cal F$}$.
$(U \hbox{\bf g})([[h]])$ as QTQHA can be transformed via the twist by 
$\hbox{$\cal F$}$ into the
quasi-triangular Hopf algebra $U_q\hbox{\bf g}$; in particular, the universal
${\cal R}$ of $U_q\hbox{\bf g}$ is given by 
$\hbox{$\cal R$}= \hbox{$\cal F$}_{21}\hbox{$\cal R$}_\Phi 
\hbox{$\cal F$}^{-1}$. }

If $H$ is the quantization of a
Poisson-Lie group associated with a solution
of the classical Yang-Baxter equation (CYBE) 
\footnote{}{In this case $H$ is is triangular,  {\it i.e.}
$\hbox{$\cal R$}_{21}\hbox{$\cal R$}_{12}={\bf 1}$}
then another 
theorem [4] states the existence of a different
$\hbox{$\cal F$}$
with similar properties as in the previous theorem---except that
now it is
enough to twist  $(U \hbox{\bf g})([[h]])$ equipped with the  ordinary
{\it coassociative} structure in order to obtain $H$. 

As shown in [5] one can
always choose $\hbox{$\cal F$}$ to be unitary, as long as $H$ is a 
compact section of  
$U_q\hbox{\bf g}$ ({\it i.e.}when $q\in {\bf R}$).
The theorems suggest that one can use the unitary twisting 
operator $\hbox{$\cal F$}$ to build
$F_{12}$ for a 2-particle sytem.
{\bf Examples:}

\item{1.} If $\hbox{$\cal A$} =\rho( U_q\hbox{\bf g})$, then we choose
$$ F =\rho^{\otimes 2}( \hbox{$\cal F$}).
$$
Ex.: $H=U_q(su(2))$, deformed quantum rotator.
\item{2.} If $\hbox{$\cal A$} = \hbox{Univ. covering Poincar\'e}
\otimes \rho(U_q\hbox{\bf g})$, were $U_q\hbox{\bf g}$ plays 
the role of an internal
symmetry, then we can set
$$ F_{12} = \hbox{id}^{(2)}_{\hbox{Poincar\'e}}\otimes 
\rho^{\otimes 2}( \hbox{$\cal F$}). 
$$
\item{3.} If $\hbox{$\cal A$}$ is the q-deformed Poincar\'e algebra of ref.
[6,7], and
$H$ is the corresponding q-deformed Lorentz Hopf algebra,
realized through $\rho$ in $\hbox{$\cal A$}$, then we can again define
$$ F_{12} = \rho^{\otimes 2}( \hbox{$\cal F$}), 
$$
where $\hbox{$\cal F$}$ belongs to the homogeneous part.
The same applies for other
inhomogeneous algebras, like the q-Euclidean ones, constructed
from the braided semi-direct
product [7] of a quantum space and of the corresponding
homogeneous quantum group. For both of these examples the
one-particle representation theory is known [13,8,9].

For $n$-particle systems one can find
$F_{12...n}$ by replacing in the previous equations
$\hbox{$\cal F$}$ by one particular element $\hbox{$\cal F$}_{12...n}$ of
$H^{\otimes n}$ satisfying the condition
$$
\Delta(x)=\hbox{$\cal F$}_{12...n}\Delta_c(x)(\hbox{$\cal F$}_{12...n})^{-1}.
\eqno{(29)}
$$%
To obtain one such a $\hbox{$\cal F$}_{12...n}$ it is enough to act on eq.
(28)  $(n-2)$ times with the coproduct in some arbitrary order.
When $n=3$, for instance, one can use
$$
\hbox{either} \hskip.9cm \hbox{$\cal F$}'_{123}:=[(\Delta\otimes id)(
\hbox{$\cal F$})]\hbox{$\cal F$}_{12}  \hskip.9cm \hbox{or}\hskip.9cm
\hbox{$\cal F$}''_{123}:=[(id\otimes\Delta)(\hbox{$\cal F$})]
\hbox{$\cal F$}_{23}.\eqno{(30)}
$$%
They coincide if $H$ is the quantization of 
a solution of the CYBE [Drinfeld]. In the the case of $U_q\hbox{\bf g}$, 
they do not coincide, but nevertheless
$\Phi:=\hbox{$\cal F$}''_{123}(\hbox{$\cal F$}'_{123})^{-1}\neq 
{\bf 1}\otimes{\bf 1}\otimes{\bf 1}$
commutes with $\Delta^{(2)}(H)$, implying that both satisfy eq.
(29). One can find a continuous family of
$\hbox{$\cal F$}_{123}$ interpolating between $\hbox{$\cal F$}'_{123}$ 
and $\hbox{$\cal F$}''_{123}$.
$(\hbox{$\cal A$}\otimes\hbox{$\cal A$}\otimes\hbox{$\cal A$})^{F_{123}}_+$ 
will depend on the specific choice
of $\hbox{$\cal F$}_{123}$; at this stage, no {\it a priori} 
preferred choice for $\hbox{$\cal F$}_{123}$ is known.

\noindent{\bf Note:}
{}From Eq. (28) follows  $(\tau\circ\Delta)(a) =
{\cal M} \Delta(a)
{\cal M}^{-1}$ with ${\cal M} := \hbox{$\cal F$}_{21} \hbox{$\cal F$}^{-1}.$
This is not the usual relation $(\tau\circ\Delta)(a) = \hbox{$\cal R$}\Delta(a)
\hbox{$\cal R$}^{-1}$ of
a quasi-triangular Hopf algebra; the latter is rather obtained by
rewriting
equation (28) in the form
$\Delta(a) = \hbox{$\cal F$} q^{t/2} \Delta_c(a) q^{-t/2} \hbox{$\cal F$}^{-1}$
where $t = \Delta_c(C_c) - 1\otimes C_c  - C_c \otimes 1$ is the invariant
tensor \
($[t , \Delta_c(a) ] = 0 \hskip.3cm\forall\hskip.3cm a \in U \hbox{\bf g}$) 
\  corresponding to
the Killing metric, and $C_c$ is the quadratic casimir of $U\hbox{\bf g}$.
 ${\cal M}$, unlike ${\cal R}$, has not nice properties
under the coproducts $\Delta\otimes id$, $id\otimes \Delta$.

The reader might wonder whether we could use  equation
$[P_{12}R, (\hbox{$\cal A$}\otimes\hbox{$\cal A$})'_+]=0$ 
(where $R=\rho^{\otimes 2}(\hbox{$\cal R$})$),
instead of eq. (6), to single out a
modified symmetric algebra $(\hbox{$\cal A$}\otimes\hbox{$\cal A$})'_+ 
\subset \hbox{$\cal A$}\otimes\hbox{$\cal A$}$; in fact,
the former is also an equation fulfilled by
$\rho^{\otimes 2}(\Delta(H))$
and reduces to the classical eq. $[P_{12}, 
(\hbox{$\cal A$}\otimes\hbox{$\cal A$})'_+]=0$ 
in the limit $q\rightarrow 1$.
However $[P_{12}R, (\hbox{$\cal A$}\otimes\hbox{$\cal A$})'_+]=0$ 
is fulfilled {\it only} by the sub-algebra
$\rho^{\otimes 2}(\Delta(H))\subset(\hbox{$\cal A$}\otimes\hbox{$\cal A$})$ 
itself, essentially because
$q^{t / 2}$ does not commute with {\it all}\/ symmetric operators,
but only with the ones corresponding to coproducts. Therefore, 
$(\hbox{$\cal A$}\otimes\hbox{$\cal A$})_+'$ defined 
via $P_{12}R$ (instead of $P_{12}^{F_{12}}$) 
is not big enough to be
in one-to-one correspondence with  the classical 
$(\hbox{$\cal A$}\otimes\hbox{$\cal A$})_+$, {\it i.e. }
is not suitable for our purposes.

Explicit universal $\hbox{$\cal F$}$'s for  
$U_q\hbox{\bf g}$ are not given in the literature; 
a $\hbox{$\cal F$}$ for a family of
deformations  (which include quantizations of solutions of
both of a CYBE and of a MCBYE) of the Heisenberg group in one
dimension was given in Ref.  [10].

However, for most practical purposes one has to deal with
representations $F$ of $\hbox{$\cal F$}$.  In
Ref. [11] a straightforward method for finding the
matrix representations of $\cal F$ from those of $\cal R$
was  proposed  and
explicit formulas were given for 
the $A,B,C,D$-series
in the fundamental representation. In [12]
matrices twisting the classical coproduct
into the q-deformed one, and therefore related to Drinfeld's twist,
were already found.

Moreover, in the intrinsic formulation of the twisted
(anti-)symmetrization
postulates [eqs. (13) -- (15)] one only needs 
the twisted
permutators $P_{12\ldots n}^{F_{12\ldots n}}$
(not the $F_{12\ldots n}$ themselves); explicit universal
expressions for the
latter may be found much more easily, as we will show
for $P_{12}^{\hbox{$\cal F$}_{12}}$ in the case $H=U_q(su(2))$.

\section{5. Explicit example: $H=U_q(su(2))$}

Let the one-particle system be a q-deformed rotator:
$\hbox{$\cal A$}\equiv \rho(H):=\rho[U_q(su(2))]$,
with $q\in {\bf R}^+$. Consider  an irreducible 
$*$-repre\-sent\-ation 
of $H$, namely 
${\cal H}\equiv V_j$, where $V_j$ denotes the highest weight 
representation of  
$U_q(su(2))$ with highest weight $j=0,{1\over 2},1, ...$. 

\vfill\eject

\section{5.1 The case of two particles}

What are $({\cal H}\otimes {\cal H})_{\pm}^{F_{12}}$ and 
$(\hbox{$\cal A$}\otimes \hbox{$\cal A$})_+^{F_{12}}$?

Let us identify $U_q(su(2))$ 
and $U(su(2))$ as algebras through the isomorphism $\phi$ 
of the first point of Drinfeld's theorem; as a consequence, $V_j$ 
carries a representation $\rho$ of both algebras. 
Similarly, $V_j\otimes V_j$ carries (irreducible) representations
of both  $U_q(su(2))\otimes U_q(su(2))$ and $U(su(2))\otimes U(su(2))$,
as well as (reducible) representations $\rho^{(2)}_c(X)$ and
$\rho^{(2)}$ of  $U_q(su(2))$ and $U(su(2))$ respectively
(by definition $\rho^{(2)}(X):=\rho^{\otimes 2}[\Delta(X)]$ and 
$\rho^{(2)}_c(X):=\rho^{\otimes 2}[\Delta_c(X)]$). Let 
$\hbox{$\cal V$}_J$ and  $\hbox{$\cal V$}^q_J$ ($0\le J\le 2j$) 
respectively denote the 
carrier spaces of the irreducible components of the latter;
from the second point of Drinfeld's theorem it follows that
$$
F_{12}\hbox{$\cal V$}_J=\hbox{$\cal V$}_J^q.\eqno{(31)} 
$$

Recall now that the $\hbox{$\cal V$}_J$'s have well-defined symmetry 
w.r.t the permutation, namely
$$
\hbox{$\cal V$}_{2j-k} \hskip.9cm \hbox{is} 
\hskip.9cm\cases{\hbox{symmetric} \cr 
\hbox{antisymmetric}\cr} 
\hskip.9cm \hbox{if k is } \hskip.9cm\cases{\hbox{even} \cr  
\hbox{odd}\cr}. \eqno{(32)} 
$$
Therefore
$$\eqalignno{
& (V_j\otimes V_j)^{F_{12}}_+:= F_{12}(V_j\otimes V_j)_+=
F_{12}\Big(\bigoplus\limits_{0\le l \le j}\hbox{$\cal V$}_{2(j-l)}\Big)=
\bigoplus\limits_{0\le l \le j}\hbox{$\cal V$}_{2(j-l)}^q  & (33) \cr
 & (V_j\otimes V_j)^{F_{12}}_-:= F_{12}(V_j\otimes V_j)_-=
F_{12}\Big( \bigoplus\limits_{0\le l\le j-{1\over 2}}\hbox{$\cal V$}_
{2(j-l)-1}\Big)=\bigoplus\limits_{0\le l\le j-{1\over 2}}\hbox{$\cal V$}
_{2(j-l)-1}^q. & (34) \cr  
}$$
We can express the content of this equation by saying 
that the subspaces $\hbox{$\cal V$}^q_J\subset V_j\otimes V_j$ have 
well-defined ``twisted symmetry''. 

Let us stress the difference between $(H\otimes H)_+^{F_{12}}$ and 
its sub-algebra $\rho^{(2)}(H)$: 
$$\eqalignno{
\rho^{(2)}(H)&\ni a:\hbox{$\cal V$}^q_J\rightarrow \hbox{$\cal V$}^q_J, & 
(35)\cr (H\otimes H)_+^{F_{12}}&\ni 
b:(V_j\otimes V_j)^{F_{12}}_{\pm}
\rightarrow (V_j\otimes V_j)^{F_{12}}_{\pm}.& (36) \cr
}$$
The elements of 
$[\rho(H)\otimes \rho(H)]_+  \setminus \rho^{(2)}(H)$ will in general 
map $\hbox{$\cal V$}^q_J$ out of itself, into some 
$\hbox{$\cal V$}^q_{J'}$'s with $J'\neq J$.

\section{5.2 Universal expression for the twisted
permutation operator of $U_q(su(2))$:}

We will say that an object
$P_{12}^{{\cal F}_{12}}$ is the ``universal  twisted
permutator'' of $U_q\hbox{\bf g}$ if 
$P_{12}^{F_{12}}=\rho^{\otimes 2} (P_{12}^{{\cal F}_{12}})$ for each
representation $\rho$ of $U_q\hbox{\bf g}$. 

In Ref. [2] we have shown that in the $\hbox{\bf g}=su(2)$ case
$$
P_{12}^{{\cal F}_{12}}=f({\bf 1}\otimes C_q)f(C_q\otimes {\bf 1})\left[f(
\Delta(C_q))\right]^{-1}\hbox{$\cal R$}_{21}\circ \tau \eqno{(37)}
$$%
Here $\tau$ is the abstract permutator ($\tau a\otimes b= b\otimes a$), 
and 
$$
C_q=X^-X^++ \left({q^{{h+1}\over 2}-q^{{-h-1}\over 2}
\over q-q^{-1}}\right)^{2} \eqno{(38)}
$$
is the casimir of $U_q(su(2))$ with 
eigenvalues $([j+{1\over 2}]_q)^2$. In the limit 
$q\rightarrow 1$: 
$C_q\rightarrow C_c+{1\over 4}$, where $C_c$ is 
the usual casimir
of $U(su(2))$ with eigenvalues $j(j+1)$. $f(z)$ is defined by
$$
\log_q[f(z)]:=\left\{{1\over \ln(q)}\sinh^{-1}\left[{(q-q^{-1})
\sqrt{z}\over 2}\right]\right\}^2-{1\over 4}; \eqno{(39)}
$$%
it is easy to verify that $f(C_q)$ has eigenvalues $q^{j(j+1)}$.

(For the proof of eq. (37), one uses the fact that the base space of
each irreducible of representation contained in a tensor
product is an eigenspace of 
$\hbox{$\cal R$}={\cal F}_{21}q^{t\over 2}{\cal F}_{12}^{-1}$ with some
eigenvalue $q^{{1\over 2}[J(J+1)-j_1(j_1+1)-j_2(j_2+1)]}$; here
$t=\Delta_c(C_c)- C_c\otimes {\bf 1} - {\bf 1}\otimes C_c$, and
$j_1,j_2=0,1,...$, $0\le J\le j_1+j_2$).

\section{5.3 The case of $n\ge 3$ particles}

When $n\ge 3$, for any given space $V$ the decomposition of 
$\bigotimes^n V$ into irreducible representations  of
the permutation group contains components with partial or 
mixed symmetry,
beside the completely symmetric and the completely 
antisymmetric ones.

As in the standard formulation, also in the twisted one
the explicit knowledge of components with mixed or partial 
symmetry  is required to build $({\cal H}^{\otimes n})_{\pm}$ 
if the Hilbert space ${\cal H}$ of one particle is the tensor product of 
different spaces corresponding to different degrees of freedom, 
${\cal H}=V\otimes V'$, as in examples 2. and 3. of section 4.

\section{6. Final remarks}

\item{1.}  We have only discussed {\it kinematics}. To discuss 
dynamics we should introduce some Hamiltonian.  Depending on 
the choice of Hamiltonian we find new physics or standard 
physics---although in a unusual picture; the latter situation
would occur e.g. if the Hamiltonian is the twist-conjugate of 
a typical Hamiltonian describing a system with an ordinary
group symmetry.

\item{2.} To generalize the above construction to
non-compact real sections of quantum groups 
(of Drinfel'd type) one should consider also
their infinite-dimensional representations and
investigate whether the corresponding
{\it formally} unitary twist $\hbox{$\cal F$}$ can be realized 
as some unitary operator even on them.

\item{3.} To introduce particle creation/annihilation operators
$A_i^{{\cal F}_{12}} ,A^{+{\cal F}_{12}}_i$ connecting twisted boson/fermion 
wave functions (second quantization) one can twist the ordinary
creation/annihilation operators $A_i,A^{+}_i$  through the
unitary operator 
$$
\hbox{$\cal F$}:= id\oplus \hbox{id}\oplus \hbox{$\cal F$}_{12} 
\oplus \hbox{$\cal F$}_{123}\oplus ...
$$ 
defined on the direct sum of all $m$-particle Hilbert
spaces, $m=0,1,2,...$.
This will lead to canonical commutation relations for
$A_i^{{\cal F}_{12}} ,A^{+{\cal F}_{12}}_i$, but these operators 
will not transform
covariantly under the quantum group action. In a separate
publication we plan to show how in the twist formalism one
can define quantum group covariant 
creation/annihilation operators; as a consequence, they
will satisfy deformed commutation relations. In the 
$U_q(su(N))$ case the result will coincide with the one
previously found in Ref. [1].

\section{\bf Acknowledgments}

We thank W.\ Pusz and S.\ L.\ Woronowicz for pointing
out reference [1] and we would like to thank J.\ Wess
for discussion and warm hospitality at his institute.

\references{Nov}{

\item{[1]} W. \spa{Pusz}, S. L. \spa{Woronowicz},
{\it Twisted Second Quantization}\/,
Reports on Mathematical Physics 27 (1989), 231-257.

\item{[2]} G.  \spa{Fiore}, P. \spa{Schupp},
{\it Identical Particles and Quantum Symmetries}\/,
Preprint LMU-TPW 95-10 (Munich University), hep-th 9508047, 
to appear in Nucl. Phys. B.

\item{[3]}
V. G. \spa{Drinfeld}, {\it Quasi Hopf Algebras}, Leningrad Math. J.
1 (1990), 1419.

\item{[4]}
V. G. \spa{Drinfeld},  Doklady AN SSSR 273 (1983) (in Russian), 531-535.
   
\item{[5]} B. \spa{Jurco}, 
{\it More on Quantum Groups from the Quantization Point of View},
Commun.\ Math.\ Phys.\ 166 (1994), 63. 

\item{[6]}
O. \spa{Ogievetsky}, W. B. \spa{Schmidke}, J. \spa{Wess} and B. \spa{Zumino},
{\it q-Deformed Poincar\'e Algebra}, Commun. Math. Phys. 150 (1992) 495-518.
   
\item{[7]}
S. \spa{Majid}, { \it Braided Momentum in the q-Poincar\'e Group},
J. Math. Phys. 34 (1993), 2045. 

\item{[8]}
M. \spa{Pillin}, W. B. \spa{Schmidke} and J. \spa{Wess}, 
{\it q-Deformed Relativistic One-Particle States}, Nucl. Phys.
B403 (1993), 223.

\item{[9]}
G. \spa{Fiore}, {\it The Euclidean Hopf algebra $U_q(e^N)$ and its
fundamental Hilbert space representations}, 
 J. Math. Phys. 36 (1995), 4363-4405;
{\it The q-Euclidean algebra $U_q(e^N)$ and the corresponding
q-Euclidean lattice}, Int. J. Mod. Phys. A, in press.

\item{[10]}
F. \spa{Bonechi}, R. \spa{Giachetti}, E. \spa{Sorace}, M. \spa{Tarlini},
{\it Deformation Quantization of the Heisenberg Group},
Commun.\ Math.\ Phys. 169  (1995), 627-633.
    
\item{[11]}
R. \spa{Engeldinger}, {\it On the Drinfel'd-Kohno Equivalence of Groups
and Quantum Groups}, Preprint LMU-TPW 95-13.
 
\item{[12]}
T. L. \spa{Curtright}, G. I. \spa{Ghandour}, C. K. \spa{Zachos},
{\it Quantum Algebra Deforming Maps, Clebsh-Gordan Coefficients, 
Coproducts, U and R Matrices},
J. Math. Phys. 32 (1991), 676-688. 
             
\item{[13]} J. \spa{Wess}, B. \spa{Zumino}, {\it Differential Calculus on
Quantum Planes and Applications}, Talk given on the occasion of the Third
Centenary Celebrations of the Mathema\-ti\-sche Gesell\-schaft Hamburg,
KA-THEP-1990-22 (1990)

}

\bye